\documentclass[sigconf]{acmart}
\usepackage{makecell}
\usepackage{svg}
\usepackage{bm}
\usepackage{tikz}
\usepackage{algorithm}
\usepackage[noend]{algpseudocode}
\usepackage{afterpage}
\usepackage{enumitem}
\usepackage{multirow}
\usepackage{multicol}
\usepackage{graphicx}
\usepackage{hyperref} 

\usepackage{lipsum}
\usepackage{tabularx}
\usepackage{array}
\usepackage{booktabs}
\usepackage{listings}

\usepackage{xcolor}
\usepackage{framed}
\usepackage{svg}
\usepackage{py}
\usepackage{rotating}
\usepackage{amsmath}
\usepackage{bm}
\usepackage{tikz}
\usepackage{placeins}
\usepackage{caption}
\usepackage{placeins}
\usepackage{float}
\newcommand{\IT}[1]{\textit{#1}}

\copyrightyear{2026}
\acmYear{2026}
\setcopyright{cc}
\setcctype{by}
\acmConference[ISLPED '26]{ACM/IEEE International Symposium on Low Power Electronics and Design}{August 05--07, 2026}{Evanston, IL, USA}
\acmBooktitle{ACM/IEEE International Symposium on Low Power Electronics and Design (ISLPED '26), August 05--07, 2026, Evanston, IL, USA}
\acmDOI{10.1145/3816440.3818667}
\acmISBN{979-8-4007-2748-1/2026/08}
\AtBeginDocument{%
  }

\ccsdesc[500]{Hardware~Power estimation and optimization}
\ccsdesc[500]{Software and its engineering~Compilers}
\ccsdesc[300]{Hardware~Memory and dense storage}

\keywords{Energy efficiency, DVFS, power gating, deep learning accelerators, non-volatile memory, compiler-directed optimization}

\begin{document}

\title{PowerFlow-DNN: Compiler-Directed Fine-Grained Power Orchestration for End-to-End Edge AI Inference}

\author
{%
  Paul Yi-Chia Chen, Jeongeun Kim, Wenbo Zhu, Yuanhan Li, Shunyao Huang, Chenjie Weng, and Christopher Torng
}

\affiliation{%
  \institution{Department of Electrical and Computer Engineering}
  \institution{University of Southern California}
  \city{Los Angeles}
  \state{CA}
  \country{USA}
}

\begin{abstract}
    Edge AI systems operate under stringent energy and volume constraints, demanding extreme efficiency on limited battery capacity, with requirements worsening as intelligent capabilities advance. Prior work suggests fine-grained power orchestration through DVFS and power gating significantly improves efficiency critical to meeting such constraints, but introduces new challenges. We observe that layer-level approaches incur unintended overheads due to inter-layer coupling of power-control decisions, and jointly managing these mechanisms under limited voltage rails and transition overheads leads to a rapidly growing combinatorial schedule space.
    We propose PowerFlow-DNN, a compiler-directed framework for end-to-end power-state orchestration in ultra-low-power accelerators. By constructing a rigorous problem formulation for deadline-constrained, real-time, periodic inference as a unified inter-layer power-scheduling problem, our framework discovers energy-minimal power-state schedules while accounting for inter-layer impacts.
    We evaluate the framework on a DNN accelerator VLSI implementation in TSMC 40nm technology. Across representative edge networks, our approach discovers near-optimal solutions and achieves energy within 0.04\% of the exact ILP oracle, reducing energy by up to 48\% compared to an aggressive baseline without power orchestration, while reasoning over a combinatorial schedule space of over $10^{160}$ possible power-state assignments, yet operating on a structured layered state graph that enables efficient optimization, achieving up to 2.14$\times$ solver speedup via lightweight pruning.
\end{abstract}

\maketitle

\section{Introduction}

Edge AI systems increasingly deploy DNN accelerators for local inference under extreme energy and volume constraints, making energy efficiency a primary design challenge due to limited battery capacities, and restricting the complexity of power delivery and voltage regulation infrastructure available on-chip~\cite{yao-TrimodalSoC-TBioCAS2021, eichler-mindful-micro2025, Sheng-epilcircuit-FP2025, Sahafi-edgewce-sr2022, abdigazy-ingestible-ne2024}.

Dynamic voltage and frequency scaling (DVFS) and power gating are widely used to improve efficiency in processors and accelerators. Most prior work assumes coarse-grained global scaling with reactive run-time control~\cite{hag-sysscale-isca2020, liu-dvfsaccel-HIPC2022}, while recent work also explores fine-grained scaling and gating with static, compiler-directed control~\cite{torng-aaws-isca2016, nayak-cgra-power-gating-trets2023}. In this work, we focus on a class of \IT{deadline-constrained, real-time, periodic inference} in embedded AI workloads that run inference at a fixed frame rate. This captures a limited but relevant subclass of edge AI with the determinism to replace expensive, predictive, or reactive dynamic power-control overheads with efficient, static, compile-time optimization of edge inference systems.

Jointly managing fine-grained power controls under practical constraints (e.g., transition overheads, limited voltage rails) remains a challenging systems-level problem. Prior work has demonstrated energy reduction through finer-grained power control that overcomes the overheads of control circuitry, using either on-chip voltage regulators~\cite{kudva-on-chip-vreg-jssc2011, seeman-on-chip-vreg-ucbphd2009} or multi-rail power supplies~\cite{miller-booster2012}.
As the computational efficiency demands of edge AI grow, and as researchers study deploying ever-smaller power domains at the macro and processing element levels~\cite{tan-icedcgra-micro2024, torng-uecgra-hpca2021, nayak-cgra-power-gating-trets2023} to capture efficiency, the power orchestration schedule space grows combinatorially with the number of power domains, voltage levels, and time interval granularity.

This work makes a key observation about the role of fine-grained power control in future edge AI systems: compiler-directed power orchestration can enable automated solvers to mitigate the challenges of scale but \textit{requires problem formulations that capture low-power VLSI implementation details} (e.g., energy and transition-time overheads, dynamic energy vs.~static leakage), so that compiler advances can guide the navigation of energy-delay tradeoffs.

To address this challenge, we propose PowerFlow-DNN (PF-DNN)\footnotemark\footnotetext{The scripts and models used in this work are available at

\url{https://github.com/usc-acorn/chen-powerflowdnn-islped2026}.}, a compiler-directed framework for power-state orchestration in ultra-low-power DNN accelerators. 
We formulate DNN inference as an inter-layer power-state scheduling problem that seeks the minimum-energy execution under a real-time deadline constraint. The challenge is not simply selecting lower voltages, but identifying that scheduled decisions at one layer impact or restrict tradeoffs in the next layer (e.g., transition and energy across different pairs of voltage rails), thereby producing a minimum-energy sequence of power states under the rail and transition constraints.

The compiler operates on a problem formulation that unifies power gating and DVFS, allowing DVFS compute islands to transition at layer boundaries (natural scheduling points), while producing power-gating schedules for power islands with high-leakage circuitry.
Finally, we show that fine-grained power orchestration decisions made by PF-DNN align well with known art and expert intuition based on the law of equi-marginal utility~\cite{azizi-eperf-analysis-isca2010, torng-aaws-isca2016} governing the fundamental policy-level tradeoff between energy (utility) and delay (cost), while allowing the compiler to solve for the detailed schedules itself.
Our results show that PF-DNN can automatically discover near-optimal schedules that achieve energy within 0.04\% of the exact oracle obtained via ILP (used as the reference solver for small instances), reducing energy by up to 48\% compared to an aggressive baseline without power orchestration and achieving up to 2.14$\times$ speedup via lightweight pruning. Our domain-aware heuristic pruning approach scales to spaces exceeding $10^{160}$, enabling efficient exploration of more complex models and architectures.

We also include a future-looking aspect that does not change the conclusions of our work but highlights its applicability: emerging non-volatile memories, specifically resistive RAM (RRAM), with bitcell densities that enable storing ML models entirely in on-chip memory banks in lieu of off-chip access~\cite{prabhu-minotaur-vlsi2024}, creating complex systems with different power profiles from the two kinds of memory.

Our primary contributions are as follows:

\begin{itemize}[noitemsep, topsep=0pt]
    \item We construct a problem formulation for a class of deadline-constrained, real-time, periodic DNN inference as a unified DVFS and power-gating control problem capturing detailed low-power VLSI constraints and cost-benefit tradeoffs.
    \item We develop a compiler-directed approach that automatically produces a near-minimal energy schedule for a problem formulation instance by integrating refinement with structured pruning to scale to a large combinatorial search space.
    \item We evaluate the complete compiler-directed power orchestration framework across representative DNN workloads on a vertically integrated toolchain spanning the compiler, architecture, and VLSI implementation in a 40nm technology.
\end{itemize}

To the best of our knowledge, we are the first to propose constructing problem formulations of fine-grained power controls with detailed VLSI constraints and allowing solvers to automatically discover efficient schedules for deterministic edge AI inference, enabling the benefits of fine-grained DVFS and power gating at scale. We hope our work lays the groundwork for future edge AI systems to exploit fine-grained power control with compiler guidance.

\section{Challenges in Edge AI Power Orchestration}
\label{sec-motivation}

\begin{figure}[t]

  \centering
  \includegraphics[width=\columnwidth]{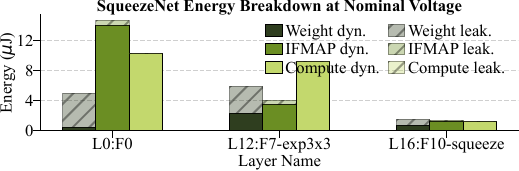}
  \vspace{-0.2in}
  \caption{Static and dynamic energy breakdown for three DNN layers, showing layer-dependent energy composition.}

  \label{fig-leakage-breakdown}
  \vspace{-0.05in}

\end{figure}


\begin{figure}[t]

  \centering
  \includegraphics[width=1\columnwidth]{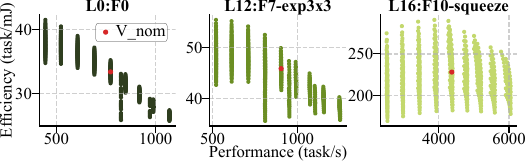}
  \vspace{-0.2in}
  \caption{Energy-performance scatter for three SqueezeNet layers under independent compute, RRAM, and feeder DVFS from 0.9V--1.2V; the red marker indicates the nominal operating point. See Section~\ref{sec-method} for methodology and scaling.}

  \label{fig-dvfs-scatter}
  \vspace{-0.05in}

\end{figure}

We first examine the workload and hardware factors that make fine-grained power orchestration in edge AI accelerators challenging.

\subsection{Energy Composition Varies Across Layers}
As edge AI workloads become more complex, the capacity of on-chip memory required to store DNN models and intermediate activations has also grown. Many accelerators, therefore, use hybrid SRAM/emerging non-volatile memory hierarchies to reduce long-term leakage~\cite{prabhu-minotaur-vlsi2024, prabhu-chimera-jssc2022, oboril-mram-tcad2015}, but focus on local energy reduction rather than coordinated system-level orchestration across layers.

Emerging memory macros often operate at lower frequencies and exhibit higher access latency than compute logic~\cite{molas-xmem-appliedsci2021, prabhu-minotaur-vlsi2024, prabhu-chimera-jssc2022}. Even with data reuse and buffering, memory access and leakage can still remain significant contributors. 
As a result, even with similar array utilization, the dominant energy source can vary across layers depending on compute intensity, data reuse, and memory access patterns, as illustrated in Figure~\ref{fig-leakage-breakdown}.
In the remainder of this paper, we use RRAM as a representative emerging memory technology.

\begin{figure*}[!t]

  \centering
  \includegraphics[width=\textwidth]{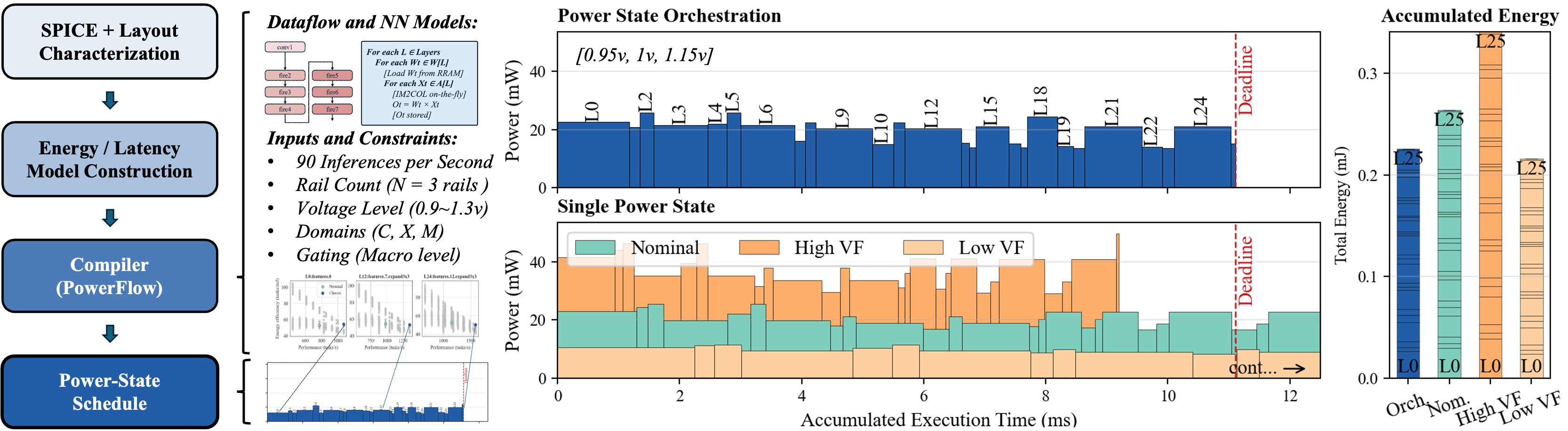}
  \vspace{-0.15in}
  \caption{PF-DNN orchestration workflow. The compiler analyzes dataflow and constraints to derive candidate power states per layer (left), then jointly schedules them across layers to meet the inference deadline while minimizing energy (right).}

  \label{fig-arch-flow}
  \vspace{-0.1in}

\end{figure*}

\subsection{Limitations of Latency-Balanced DVFS}
DVFS is a widely used technique for improving energy efficiency in computing systems~\cite{Geng-powerlens-DAC2026, hag-sysscale-isca2020, isci-dvfs-micro2006, weiser-dvfs-osdi1994, eyerman-fgdvfs-taco2008, zhai-udvfs-tvlsi2005, torng-aaws-isca2016}. Recent works have explored predictive and adaptive DVFS policies for hardware accelerators and heterogeneous systems~\cite{hag-sysscale-isca2020, liu-dvfsaccel-HIPC2022}, but do not capture inter-layer dependencies introduced by shared voltage rails.

In contrast to general-purpose processors or accelerators, many embedded and edge accelerators execute deterministic inference workloads with fixed computation graphs~\cite{Sahafi-edgewce-sr2022, abdigazy-ingestible-ne2024, tan-icedcgra-micro2024}. In such settings, dynamic run-time DVFS policies with heuristics introduce unnecessary control complexity and may incur excessive switching overhead, leading to suboptimal energy efficiency.

Figure~\ref{fig-dvfs-scatter} illustrates the energy–latency tradeoffs obtained by sweeping the voltages of compute and memory domains for several representative layers. The minimum-energy operating point varies across layers and deadlines with different voltage combinations, indicating that a single global DVFS policy cannot capture the true energy-optimal schedules under deadline and rail constraints.

\subsection{Impact of Rail Scarcity Constraint}
While fine-grained voltage scaling has been explored in modern systems~\cite{torng-uecgra-hpca2021,godycki-rpdn-micro2014, liu-dvfsaccel-HIPC2022, miller-booster2012, truong-167-processor-jssc2009, tan-icedcgra-micro2024}, practical designs often expose limited supply rails due to power-delivery and integration constraints. This is particularly evident in extreme-edge platforms such as biomedical devices~\cite{abdigazy-ingestible-ne2024, abdigazy-gas-cr2024, yao-TrimodalSoC-TBioCAS2021}, where form-factor limitations limit voltage regulation. 
As a result, voltage levels must be shared across multiple domains rather than independently controlled. This shared-rail constraint prevents fully independent scaling of compute, memory, and data-movement domains, introducing a system-level challenge: selecting and coordinating a small set of voltage rails across layers while accounting for transition overheads.

\subsection{Granularity vs Orchestration Tradeoff}
Increasing the granularity of power control (additional voltage rails or independent power domains) enables more precise matching between hardware resources and workload demands. However, this flexibility also grows the orchestration design space combinatorially as the number of layers, domains, and levels increases. Consequently, naive heuristics or manual tuning become ineffective, motivating structured optimization to efficiently explore this space.
Effective energy minimization requires coordinated power state orchestration across layers. Recent work has explored inter-layer and spatial orchestration in accelerators and reconfigurable architectures~\cite{gong-crane-micro2025, bai-klotskiv2-tcad2024, torng-uecgra-hpca2021, tan-icedcgra-micro2024, gobieski-riptide-micro2022, serafin-pipestitch-micro23, gobieski-snafu-isca2021}. While these approaches coordinate computation and mapping across layers, space, and time, they do not explicitly address power management as a unified optimization problem across DVFS, power gating, and voltage levels.

\section{Architecture and Framework Overview}
\label{sec-arch-framework}
PF-DNN targets accelerators composed of multiple voltage domains that share limited supply rails.
Figure~\ref{fig-arch-flow} shows the overall orchestration workflow with a demonstrated schedule.

\subsection{Target Architecture Model}
We model the accelerator as a set of controllable power-managed units
$D=\{D_1,\dots,D_K\}$, where each unit may correspond to a coarse DVFS-controlled domain (e.g., compute, feeder, and the RRAM memory subsystem) or a finer-grained gated memory unit. Each unit operates either at a selected voltage rail or in an active/gated state, depending on its control granularity.
Inference executes as a sequence of operations derived from the neural-network dataflow graph. In our prototype system, these operations correspond to neural-network layers, although the formulation applies more generally to any sequence of computational phases.
Though our evaluation focuses on a systolic-array accelerator with SRAM buffers and RRAM-based weight storage, the framework applies to any accelerator that exposes multiple voltage domains and power-gating capabilities.

\subsection{Power States and Scheduling Anchors}
PF-DNN models DVFS and power gating as unified power-state scheduling decisions. As a result, tradeoffs between reducing dynamic energy (via DVFS) and leakage (via gating) can be evaluated jointly.
For operation $i$, a valid operating state specifies the voltage assignments of DVFS-controlled units and the gating schedules of memory-controlled units. These operating states are later formalized as decision variables in the optimization problem.

\label{sec-scheduling-anchors}
Scheduling decisions occur at execution boundaries (anchors).
Layer boundaries provide natural coarse-grained anchors because each exhibits stable computation and memory characteristics. Voltages assigned at layer boundaries avoid the overheads of frequent intra-layer rail switching.
At a finer granularity, memory-access phases provide additional anchors for macro gating. The banks with unused weights during portions of execution can be temporarily gated to reduce leakage, identified by compiler analysis of layer dataflow and access patterns.
Consequently, PF-DNN performs inter-layer scheduling of voltage assignments while enabling intra-layer memory gating decisions derived from compiler analysis.

\subsection{Compiler Workflow}
\label{sec-compiler-workflow}
Compilation occurs once per deployment. The compiler first analyzes the workload dataflow graph to determine memory occupancy, data movement, and domain activity across phases, producing a set of feasible operating points for each operation, each with a distinct combination of power states.
Given these candidates, PF-DNN enumerates candidate rail subsets and determines the minimum-energy feasible schedule under each subset, selecting the overall best solution that satisfies the target inference deadline, as shown in Figure~\ref{fig-arch-flow}. The resulting voltage assignments and memory-gating decisions are compiled and programmed into the on-chip memory as a static schedule, along with the layer definitions used during run-time execution, while the $pg\_manager$ manages the inter-layer fine-grained memory-gating schedules. 

\section{Problem Formulation}

Based on the architecture model and compiler, we formulate power-state orchestration as a constrained optimization problem.

\subsection{System Model}
A neural network inference consists of $L$ sequential operations (layers). The accelerator operates on a subset of available voltage rails $\mathcal{R} \subseteq \mathcal{V}$, where $\mathcal{V}$ denotes the discretized set of candidate voltage levels. For each layer $i$, the solver selects a state $s_i \in \mathcal{S}_i(\mathcal{R})$, where each state specifies a domain-wise assignment $\{V_i^{D_0}, \ldots, V_i^{D_K}\}$, with $V_i^{D_k} \in \mathcal{R} \cup \{0\}$ and $0$ denoting a gated state. The set $\mathcal{S}_i(\mathcal{R})$ contains all valid operating points for layer $i$ under $\mathcal{R}$. Each state $s_i$ is characterized by an execution latency $T_{\text{op}}(s_i)$ and an energy $E_{\text{op}}(s_i)$, capturing both static and dynamic components.

Switching between consecutive states incurs latency and energy overhead, captured by transition functions $T_{\text{trans}}(s_{i-1}, s_i)$ and $E_{\text{trans}}(s_{i-1}, s_i)$. Transitions are assumed not to overlap with computation. The accelerator supports at most $N_{\max}$ voltage rails. For deterministic edge inference workloads, the inference deadline is a fixed scalar: $T_{\max} = 1/R_{\text{target}}$. For completeness, we include a terminal state $s_{L+1}$ to model the idle interval before $T_{\max}$.

\subsection{Optimization Objective}
\label{sec-opt-obj}
The continuous voltage range $[V_{\min}, V_{\max}]$ is discretized into a finite candidate set $\mathcal{V}$ using a uniform step size $\Delta V$. The optimization selects a rail subset $\mathcal{R} = \{V_1, \dots, V_N\}$, where $V_n \in \mathcal{V}$ and $|\mathcal{R}| \le N_{\max}$, and selects a sequence of states $s_i \in \mathcal{S}_i(\mathcal{R})$.

Total inference time is
\begin{equation}
T_{\text{infer}} =
\sum_{i=1}^{L} T_{\text{op}}(s_i)
+ \sum_{i=1}^{L} T_{\text{trans}}(s_{i}, s_{i+1})
\end{equation}
where transition overheads capture rail-switching and memory wake events.
We formulate cross-layer power-state orchestration as the following optimization:
\begin{equation}
\min_{\mathcal{R},\, s_1,\dots,s_L,\, s_{L+1}}
E_{tot} = 
\sum_{i=1}^{L} E_{\text{op}}(s_i) 
+ \sum_{i=1}^{L} E_{\text{trans}}(s_{i}, s_{i+1}) 
+ E_{\text{idle}}(s_{L+1}) 
\end{equation}
\begin{equation}
\text{s.t.} \;
|\mathcal{R}| \le N_{\max}, \;
\mathcal{R} \subseteq \mathcal{V}, \;
T_{\text{infer}} \le T_{\max}, \;
s_i \in \mathcal{S}_i(\mathcal{R}), \;
i = 1,\dots,L.
\nonumber
\end{equation}
The terminal state $s_{L+1}$ models the idle interval following the final transition. Let $z \in \{0,1\}$ denote the duty-cycling decision, where $z=1$ indicates that the accelerator remains active during the idle interval. The idle energy is given by $E_{\text{idle}}(s_{L+1}) = z \cdot P_{\text{idle}} \cdot [T_{\text{max}}-T_{\text{infer}}]$, where $P_{\text{idle}}$ denotes the idle power, capturing both static leakage and residual dynamic power under clock gating. Thus, idle energy is incurred only when the accelerator remains active.

One-time configuration energy is excluded because it is independent of the selected schedule. Voltage-frequency behavior is captured in each state's latency and energy. The resulting search space grows combinatorially with rail subsets, layers, domains, and voltage levels, making exhaustive search intractable.

Transition costs are modeled as pairwise functions between states, capturing asymmetric and domain-dependent switching behavior. Although our evaluation uses a simplified model, the formulation supports detailed characterization-driven transition costs without modification.

From the defined optimization function, the full schedule space is upper-bounded by $\sum_{k=1}^{N_{\max}} \binom{|\mathcal{V}|}{k} (k+1)^{DL}$, which grows exponentially in the number of layers and domains. While this defines a worst-case combinatorial schedule space, the DP-based algorithm instead operates on a layered state graph of feasible per-layer states. The effective complexity therefore depends on the number of states across layers, $\sum_{i=1}^{L+1} |\mathcal{S}_i|$, and the transitions between adjacent layers, $\sum_{i=1}^{L} |\mathcal{S}_i||\mathcal{S}_{i+1}|$, rather than the full schedule space.

\subsection{Solution Approach}
While ILP provides the exact solution for the constrained formulation, it becomes impractical for large full-network instances due to combinatorial growth in the scheduling space. Consequently, we rely on a dynamic-programming (DP) algorithm for scalable optimization. We use ILP only for validation on small instances.

For each operation, the framework enumerates feasible operating states at different voltage levels with associated latency and energy, as described in Section~\ref{sec-compiler-workflow}. To incorporate layerwise coupling, PF-DNN models the scheduling as a shortest-path problem over a layered state graph under rail constraints. We solve the deadline-constrained problem using a Lagrangian DP-based search ($\lambda$-DP), where $\lambda$ reweights the objective as $E + \lambda T$. Since this weighted search may miss lower-energy feasible schedules that no $\lambda$ can represent, we add a local refinement step: among up to ten feasible candidate paths, the compiler greedily applies up to eight single-layer replacement moves, each chosen from all layers and accepted only if it reduces total energy while preserving the timing deadline and selected rail constraint.
To improve solver run time, we also evaluate \textit{structure pruning}, which removes locally dominated states within each layer to reduce the space to be explored.
Algorithm scalability and validation are evaluated in Section~\ref{sec-eval-solver}.

\section{Evaluation Methodology}
\label{sec-method}

\begin{figure}[t]

  \centering
  \includegraphics[width=\columnwidth]{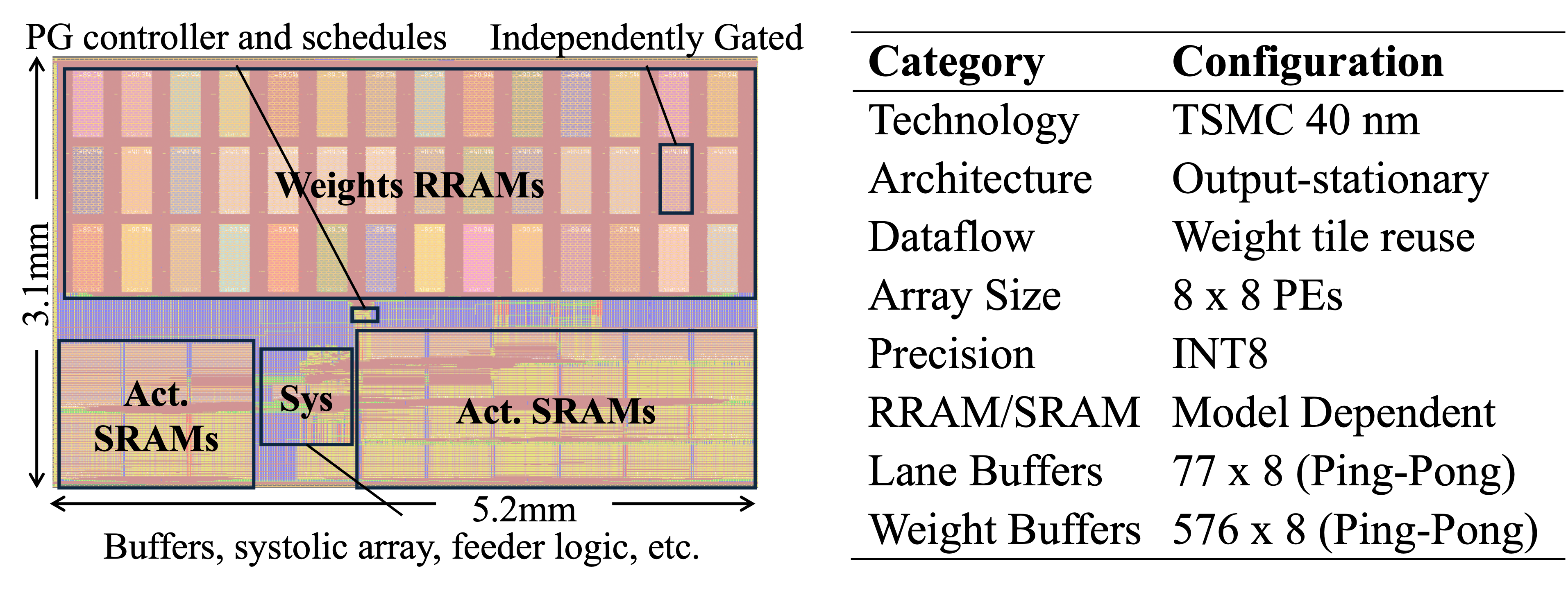}
  \begingroup
  \captionsetup{skip=1pt}
  \caption{40nm accelerator layout. The chip operates up to 500~MHz, with the RRAM subsystem at 100~MHz.}


  \label{fig-layout}
  \vspace{-0.1in}
  \endgroup

\end{figure}

To study the proposed co-optimization problem, we model the accelerator subsystem under a limited rail-count constraint. 
\subsection{Hardware and Modeling Setup}
Table in Figure~\ref{fig-layout} summarizes our evaluation parameters. We synthesize and place-and-route a representative programmable accelerator instance in TSMC 40nm LP technology, shown in Figure~\ref{fig-layout}, to obtain post-layout timing, area, and energy for characterization. The physical instance uses a SqueezeNet-sized memory bank and buffer configuration for calibration, while the number of memory banks, layer definitions, and dataflow parameters are model-dependent, and power schedules are generated by the compiler.

RRAM bank activity is determined from the deterministic weight-address stream generated by the DMA engine, which reflects the accelerator's dataflow schedule. These events serve as scheduling anchors as mentioned in Section~\ref{sec-scheduling-anchors}. The run time follows the compiler-generated schedule without additional control logic.
Energy consumption is obtained from post-synthesis gate-level power analysis and used to construct a per-event energy lookup model capturing compute, memory, and data-movement events. SRAM access energy and leakage are extracted from the TSMC memory compiler, while RRAM characteristics follow prior work~\cite{prabhu-minotaur-vlsi2024, prabhu-chimera-jssc2022}. All workloads use INT8 precision for both weights and activations.

A cycle-accurate performance model is used to estimate layer latency and generate activity counts for the energy model, validated against RTL simulations. Candidate schedules are validated against post-layout timing limits obtained from static timing analysis at the typical PVT corner. Energy is reported per interval in the continuous inference setting, where idle periods between inferences depend on the target inference rate and compiler schedule.

In practice, the target PVT corner should match the deployment environment to remain timing-safe under real-world conditions.

\subsection{DVFS and Transition Modeling}

Voltage-frequency scaling is derived from SPICE characterization of an FO4-loaded ring oscillator in the target 40nm LP process. Energy scaling follows a first-order voltage–frequency model~\cite{torng-uecgra-hpca2021}. Voltages range from 0.9–1.3 V (step 0.05).

We assume schedule updates are prepared in parallel via shadow configuration registers at scheduling boundaries. As a result, configuration latency is treated as a one-time offset and is not modeled as part of the execution or transition latency.

Switching between supply rails incurs latency, energy, and area overhead due to charging and discharging the domain capacitance through the power-switch network. Consistent with prior studies~\cite{usami-fgpg-vlsid2009, xue-regate-micro2025}, we assume worst-case transition latencies from circuit-level modeling: 5~ns for memory wake-up and 15~ns for DVFS rail switching.
Transition energy is modeled as $E_{switch} = C_{dom}(V_{high}^2 - V_{low}^2)$. We assume a nominal transition energy of 1~nJ and evaluate sensitivity over 0.1~nJ–1~$\mu$J. While level shifter overheads between voltage domains are not explicitly modeled, their impact is small relative to memory and compute energy, and can be incorporated into the transition cost model without changing the formulation.

We capture area overhead at the chip level, including the centralized \textit{pg\_manager} and three clock-generation blocks, which contribute 0.24\% of total area. For larger domain schedules, domain-level overhead is extrapolated from per-macro measurements, resulting in 4.6--9.2\% area across domains depending on domain size.

\subsection{Workloads and Metrics}
We evaluate four representative edge networks: SqueezeNet1.1 (26 layers, Conv/Fire)~\cite{Iandola-squeeze-arx2016}, MobileNetV3-Small (52 layers, DW/Conv/SE)~\cite{Howard-mobilenetv3-ICCV2019}, ResNet18 (20 layers, Conv/Residual)~\cite{he-resnet-iccv2016}, and MobileViT-xxs (72 layers, Conv/Attention)~\cite{Mehta-mobilevit-arx2021}. These workloads span conv, depthwise, residual, and attention layers, producing diverse dataflows and power-management opportunities.

\section{Evaluation}
\label{sec-eval-baseline}
We evaluate whether PF-DNN identifies the minimum-energy operating point under discretized voltage and power-state spaces, and compare against strong heuristic baselines. Specifically, we examine:
(1) energy reduction across inference rates compared to baseline policies,
(2) generalization across representative networks,
(3) the impact of hardware constraints, and
(4) solver scalability. 

The baseline corresponds to a conventional accelerator design without cross-layer power optimization~\cite{chen-eyeriss-JSSC2016}. The +gating variant extends this design with fine-grained memory power gating inspired by prior work~\cite{prabhu-minotaur-vlsi2024, prabhu-chimera-jssc2022}. The +greedy policy implements a marginal-utility-based layer-wise DVFS heuristic, inspired by prior accelerator DVFS approaches~\cite{liu-dvfsaccel-HIPC2022,Geng-powerlens-DAC2026,torng-uecgra-hpca2021}. Starting from the minimum-energy configuration, the heuristic iteratively applies per-layer voltage adjustments that provide the largest latency reduction per unit energy increase until the target deadline is met. While transition overheads are considered during candidate evaluation, decisions are made locally and independently. We compare these baselines, their combination, and the proposed PF-DNN under identical hardware and timing constraints.

\subsection{Energy vs Inference Rate}

\begin{figure}[t]

  \centering
  \includegraphics[width=1\columnwidth]{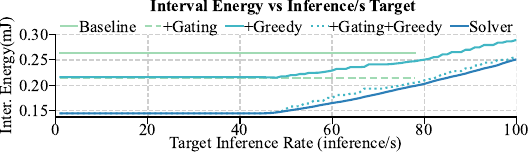}
  \begingroup
  \vspace{-0.2in}

  \caption{Inference interval energy vs. target inference rate for SqueezeNet. We compare baseline, +gating, +greedy, +gating+greedy, and the proposed method with rail selection (see Section~\ref{sec-eval-baseline} for definitions).}

  \label{fig-eval-energy-vs-rate}
  \vspace{-0.05in}
  \endgroup
\end{figure}
Figure~\ref{fig-eval-energy-vs-rate} shows inference interval energy sweeping the target inference rate. Fine-grained power gating significantly reduces energy by eliminating leakage from idle RRAM banks. In contrast, +greedy considers transitions only locally and cannot coordinate power-state assignments across layers. Combining DVFS and power gating improves energy efficiency and is sufficient at lower inference rates where slack is abundant. However, it remains suboptimal at higher rates due to locally optimized decisions. PF-DNN achieves lower energy by jointly optimizing voltage rail selection and power-state schedules while accounting for transition overheads.

\subsection{Generalization Across Models}

\begin{figure}[t]

  \centering
  \includegraphics[width=1\columnwidth]{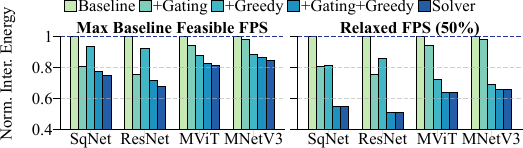}
  \begingroup
  \captionsetup{skip=1pt}

  \caption{Normalized inference interval energy across four edge models under tight and relaxed deadlines.}

  \label{fig-cross-model}
  \vspace{-0.05in}
  \endgroup
\end{figure}
To evaluate generality, we repeat the experiment across four representative edge networks, as shown in Figure~\ref{fig-cross-model}. At each model’s maximum feasible inference rate, PF-DNN reduces energy by 34--48\% relative to the unoptimized baseline and further improves up to ~5\% over the +greedy+gating baseline. While the latter gain is modest, it is achieved over an aggressive layer-wise heuristic and remains meaningful in edge deployments. Under relaxed deadlines, the two approaches converge because abundant slack allows layers to operate near their minimum-energy points.

\subsection{Impact of Voltage Rail Availability}

\begin{figure}[t]

  \centering
  \includegraphics[width=1\columnwidth]{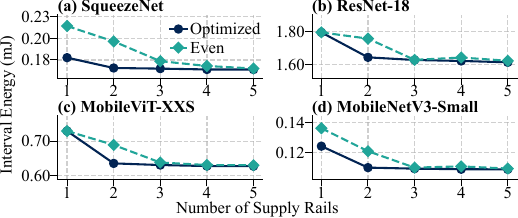}
  \begingroup
  \captionsetup{skip=2pt}
  \caption{Inference interval energy vs. voltage rail count. We compare evenly spaced rails and optimized rail selections, both evaluated under PF-DNN orchestration.}

  \label{fig-energy-vs-rail-count}
  \vspace{-0.1in}
  \endgroup

\end{figure}
Voltage rail availability directly affects the achievable minimum energy. As shown in Figure~\ref{fig-energy-vs-rail-count}, increasing the number of rails allows the scheduler to reach a lower energy operating point. Energy decreases by 7.7–14\% when increasing from one to three rails, with diminishing returns beyond three rails. Additionally, jointly optimized rail level selection improves efficiency by up to 17\% compared to evenly spaced rails when the number of rails is limited. This suggests that a small number of carefully chosen rails captures most of the achievable energy savings, highlighting that rail selection is more impactful than increasing the number of voltage levels.

\subsection{Additional Observations}

\begin{figure}[t]

  \centering
  \includegraphics[width=1\columnwidth]{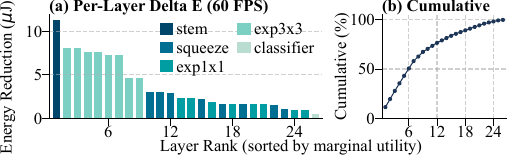}
  \begingroup
  \captionsetup{skip=2pt}
  \caption{Layers are ranked by local marginal utility, defined as energy reduction per unit latency increase from the nominal operating point. Bars show per-layer energy reduction for the compiler output schedule as Figure~\ref{fig-arch-flow}.}

  \label{fig-marginal-utility}
  \vspace{-0.05in}
  \endgroup

\end{figure}

The orchestration framework enables analysis to surface insights into different hardware choices. Separating compute and memory into independent domains extended the energy reduction by 11\%, while further partitioning provided diminishing returns. Fine-grained memory gating reduces leakage by up to 90\%, explaining why the greedy baseline achieves similar results in some relaxed-deadline regimes. When $E_{\text{trans}}$ is swept across four orders of magnitude (0.1~nJ-1~$\mu$J), PF-DNN suppresses switching, reducing rail-transition counts by up to 97\% (e.g., 74 to 2 for MobileNet). Energy savings are skewed across layers, with most reduction coming from a small subset (least utility, highest cost) as shown in Figure~\ref{fig-marginal-utility}, aligning with known intuition on the law of equi-marginal utility~\cite{azizi-eperf-analysis-isca2010, torng-aaws-isca2016}.

\subsection{Solver Scalability and Variants}
\label{sec-eval-solver}

\begin{figure}[t]

  \centering
  \includegraphics[width=1\columnwidth]{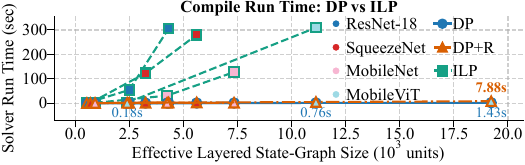 }
  \begingroup
  \captionsetup{skip=2pt}
  \caption{Solver run time versus explored layered-state graph size (feasible state graph defined in Section~\ref{sec-opt-obj}).}
  \Description{}
  \label{fig-eval-compiler}
  \endgroup
  \vspace{-0.1in}
\end{figure}
Finally, we evaluate solver scalability in Figure~\ref{fig-eval-compiler} with a smaller voltage step to remain tractable. Though run time scales with the layered state graph rather than the full combinatorial schedule space, the latter helps explain the benefits of the proposed DP and why ILP fails. The results show that ILP serves as an oracle but scales poorly because it instantiates binary variables and transition constraints over layer-state pairs. As the number of feasible states and adjacent-layer transitions grows, the ILP runs out of memory even with reduced voltage levels. In contrast, $\lambda$-DP operates directly on the layered state graph and stores only the DP frontier, returning full-network solutions within seconds, while the $\lambda$-DP with refinement takes around three to six times longer, reducing the gap with the ILP oracle from 1.43\% to 0.04\% of the optimum.
\label{sec-solver-variants}
Also, structure pruning produces identical schedules to the unoptimized solver while improving run time by up to 2.14$\times$.

Across all experiments, PF-DNN consistently identifies near-minimum-energy schedules, showing that a small number of rails captures most energy savings, compute/memory domain separation improves efficiency, and orchestration naturally suppresses excessive switching as transition costs increase.


\section{Conclusion}
Fine-grained DVFS and power gating are increasingly available in modern AI accelerators, but exploiting this flexibility is challenging in rail-constrained systems with shared voltage domains. Our results show that layer-wise DVFS heuristics fail to capture inter-layer transition costs and shared-rail interactions. By formulating power management as a unified orchestration problem, PF-DNN identifies energy-efficient schedules that satisfy inference deadlines under realistic constraints for deterministic workloads. These results demonstrate that fine-grained control alone is insufficient. Structured orchestration is required to realize its full potential.
Future work includes extending PF-DNN to jointly optimize accelerator mapping, scheduling, and resource allocation. Such co-optimization may further improve energy efficiency in systems with dynamic workloads, shared resources, and heterogeneous accelerators.

\paragraph{Acknowledgment and Disclosure.}
The authors used OpenAI ChatGPT to assist in developing plotting and visualization templates.
\bibliographystyle{ACM-Reference-Format}
\bibliography{refs}
\end{document}